\documentclass[12pt,a4paper]{article}

\usepackage{epsf}

\newcommand{\beq}{\begin{equation}}
\newcommand{\eeq}{\end{equation}}
\newcommand{\beqa}{\begin{eqnarray}}
\newcommand{\eeqa}{\end{eqnarray}}

\newcommand{\vs}{\vspace{-0.25cm}} 
\newcommand{\Mp}{M_{\pi^\pm}}
\newcommand{\MK}{M_{K^\pm}}
\newcommand{\Mn}{M_{\pi^0}}
\newcommand{\MKn}{M_{K^0}}
\newcommand{\qin}{{\bf q_{\rm in}}}
\newcommand{\qout}{{\bf q_{\rm out}}}

\begin{document}

\hfill {\tiny FZJ-IKP(TH)-2001-11}  
  
\bigskip\bigskip\bigskip 
 
\begin{center} 

{{\Large\bf Isospin violation in 
   pion--kaon scattering\footnote{Work supported in part
   by funds provided by the ``Studienstiftung des deutschen Volkes''.}
}} 
 
\end{center} 
 
\vspace{.2in} 
 
\begin{center} 
{\large  Bastian Kubis\footnote{E-mail: b.kubis@fz-juelich.de},
 Ulf-G. Mei{\ss}ner\footnote{E-mail: Ulf-G.Meissner@fz-juelich.de}}

\bigskip 
 
\bigskip 
 
{\it Forschungszentrum J\"ulich,  
Institut f\"ur Kernphysik (Theorie)\\  
D--52425 J\"ulich, Germany}

\end{center} 
 
\vspace{.7in} 
 
\thispagestyle{empty}  
 
\begin{abstract} 
\noindent 
We consider strong and electromagnetic isospin violation in 
near--threshold pion--kaon scattering. At tree level, such effects are
small for all physical channels. We work out the complete one--loop
corrections to the process $\pi^- K^+ \to \pi^0 K^0$. They come out
rather small. We also show that the corresponding
radiative cross section is highly suppressed at threshold.
\end{abstract} 
 
\vspace{1.3in} 
 
 
\centerline{Keywords:  
{\it pion--kaon scattering}, {\it electromagnetic corrections}, {\it chiral perturbation theory}}
 
\vfill

\newpage 
 

\section{Introduction}

\noindent The quark masses allow to consider various approximations
to Quantum Chromodynamics (QCD). For the $c$, $b$, and $t$ quarks, the masses
are so large that an expansion in inverse powers of these masses can be
performed systematically. This leads to the so--called heavy--quark effective field theory. 
On the other hand, to a good first approximation, one can 
consider the light quarks $u$, $d$, and $s$ as massless. In that limit, the
QCD Lagrangian exhibits a chiral symmetry which is, however,
spontaneously broken as witnessed by the 
appearance of eight almost massless pseudoscalar Goldstone bosons,
the pions, the kaons, and the eta. These would--be Goldstone bosons
acquire their masses from the explicit symmetry breaking due to the
quark mass term. Spontaneous as well as explicit symmetry violation
can be explored systematically by means of 
chiral perturbation theory (ChPT), an effective field theory
formulated in terms of the asymptotically observable fields. 
At low energies, the generating functional of
QCD is characterized by two energy scales. 
One is the pion (kaon) decay constant in the chiral limit, denoted by $F$. 
Its non-vanishing value ($F \simeq 88$~MeV) is a necessary and sufficient condition
for spontaneous symmetry breaking, much like the vacuum expectation
value of the neutral Higgs boson signals the breaking of the
electroweak symmetry. 
The second scale is  given by the quark
condensate, $B = |\langle 0 | \bar{q}q|0\rangle| / F^2$.
In the standard scenario of chiral symmetry breaking, 
$\langle 0|\bar{q}q|0 \rangle \simeq (-225~{\rm MeV})^3$ so that 
$B \simeq 1.5$~GeV$\, \gg F$. 
This leads e.g.\ to a very precise prediction for the
S--wave $\pi \pi$ scattering lengths~\cite{CGL}. This scenario seems to be
confirmed by recent Brookhaven data on $K_{\ell 4}$ 
decays \cite{BNL} and will be
further scrutinized when the pionium lifetime measurements performed
at CERN~\cite{DIRACpipi} have been analyzed. 
For the three flavor sector, the situation is, however, less clear. 
Indeed, the observation that $m_s \simeq
\Lambda_{\rm QCD}$ has even led to investigations considering the
strange quark as heavy \cite{CK,Roessl,Ou}. Furthermore, there have been recent
speculations that the structure of the QCD vacuum changes dramatically
with increasing number of flavors, e.g.\ it could be possible that
the condensate is sizeably suppressed in SU(3) as compared to SU(2) \cite{Descotes}.

\medskip \noindent
One of the cleanest processes to test our understanding of the
symmetry breaking pattern in the presence of strange quarks is
elastic pion--kaon ($\pi K$) scattering near threshold. This reaction
is interesting for a variety of reasons. First, it is very similar to 
$\pi \pi$ scattering in the two--flavor sector but also different
in that the quark mass difference $m_u \!-\! m_d$ can appear at leading
order in the $\pi K$ scattering amplitude. Second, there exist
abundant data from inelastic processes which allow one to extract 
low--energy characteristics of the $\pi K$ scattering amplitude by means of
dispersion theory. The existing determinations of the S--wave
scattering lengths are, however, plagued by large uncertainties, 
see e.g.\ 
\cite{BKMpiK1}. For  recent work on combining 
dispersion relations and chiral perturbation theory, see \cite{AB}.
Third, the DIRAC collaboration 
intends to measure the lifetime of $\pi K$ atoms at CERN
\cite{DIRACpiK} which gives 
direct access to the isovector S--wave scattering length. 
To also pin down the isoscalar S--wave scattering length, 
one would have to measure the $2P-2S$ level shift,
similar to what has been done for the pion--nucleon system at PSI \cite{PSI}.
The precise relation  between the $\pi K$ atom lifetime and the
scattering length (the so--called modified Deser formula) can be
worked out by means of an effective field theory for hadronic bound
states (for the pionium case see e.g.~\cite{GGLR,Orsay}). 

\medskip \noindent 
To achieve the necessary accuracy in the $\pi K$ system, it is
mandatory to sharpen the existing one--loop predictions for the
scattering lengths and range parameters \cite{BKMpiK1,BKM} by including
isospin violation due to strong and electromagnetic effects. This is done here.
First, we study isospin violation for the S--wave scattering lengths at tree level for all physical
channels. This allows for a first estimate of such effects and is
also interesting to compare directly to the $\pi\pi$ case. Then we
focus on the one--loop strong and electromagnetic corrections to the
relevant channel for $\pi K$ atoms, $\pi^- K^+ \to \pi^0 K^0$,
including also a complete treatment of soft photon radiation, 
$\pi^- K^+ \to \pi^0 K^0 \gamma$.


\section{Lagrangians}

\noindent
In this section we discuss the strong and electromagnetic effective Lagrangians
underlying our calculations. All the pieces have been discussed extensively
elsewhere in the literature, so we will just present the terms needed for
the following. The effective Lagrangian can be expanded at low energies according to
\beq
{\cal L}_{\rm eff} = {\cal L}^{(2)}_{\rm str} + {\cal L}^{(2)}_{\rm em}
+ {\cal L}^{(4)}_{\rm str} + {\cal L}^{(4)}_{\rm em} + \ldots ~,
\eeq
where the superscripts (2), (4) refer to the chiral dimension and ``str'' and ``em''
denote the strong and electromagnetic terms, respectively. 
Chiral power counting for the strong sector attributes the chiral dimension $q$
to all pseudo--Goldstone boson masses ($M_\pi$, $M_K$, $M_\eta$) as well as all
momenta involved. 
This scheme is generalized to include electromagnetic terms by counting 
the electric charge $e$ also as a quantity of order $q$, which is
dictated by the requirement that a unique dimension be assigned to the covariant
derivative (that includes the lowest order photon coupling). 
Therefore any term of the form $e^{2j} q^{2k} M_\pi^{l} M_K^{m} M_\eta^{n}$
with, e.g.,  $2j+2k+l+m+n=4$ is counted as fourth order.

\medskip \noindent 
The lowest--order Lagrangian is given by
\beqa
{\cal L}^{(2)} &=& - \frac{1}{4} F_{\mu\nu}F^{\mu\nu} 
               - \frac{\lambda}{2}(\partial_\mu A^\mu)^2 \nonumber\\
     &+&  \frac{F^2}{4} \langle D_\mu U^\dagger D^\mu U 
        + \chi U^\dagger + \chi^\dagger U \rangle
        + C \langle Q U Q U^\dagger \rangle ~. 
\label{L2}
\eeqa
Here, $\langle \ldots \rangle$ denotes the trace in flavor space.
$F_{\mu\nu}=\partial_\mu A_\nu - \partial_\nu A_\mu$ is the usual electromagnetic
field strength tensor, $\lambda$ refers to the gauge fixing parameter. 
All calculations were performed in the Feynman gauge, $\lambda=1$.
$U=\exp (i\Phi/F)$ collects the (pseudo--)Goldstone boson fields,
the low--energy constant (LEC) $F$ can be identified with a common
meson decay constant in the chiral limit.
$D_\mu U = \partial_\mu U -i[v_\mu,U]-i\{a_\mu,U\}$ defines
the covariant derivative acting on $U$ in the presence of external
vector ($v_\mu$) and axial vector ($a_\mu$) currents. $\chi$ includes
external scalar ($s$) and pseudoscalar ($p$) sources, $\chi = 2B(s+ip)$,
where for our purposes, only the quark mass term in the source $s$
is of relevance, $s={\cal M} + \ldots\,$, ${\cal M} = {\rm diag}(m_u,m_d,m_s)$.
The LEC $B$ is linked to the quark condensate at leading order as already
discussed in the introduction.
$Q$ is the quark charge matrix, $Q = e\,{\rm diag}(2/3,-1/3,-1/3)$.
The LEC $C$ accompanying the last term in the Lagrangian eq.~(\ref{L2}) 
can be calculated from the leading order electromagnetic pion mass difference,
$\Mp^2\!-\!\Mn^2 = 2Ze^2F^2$ (neglecting a tiny strong mass difference $\sim(m_u-m_d)^2)$, 
where we have defined the convenient dimensionless constant $Z=C/F^4$. Using $F=F_\pi=92.4$~MeV,
one obtains $Z\approx 0.8$.

\medskip \noindent
The fourth--order Lagrangian \cite{NPB250} contains the following ``strong'' terms
which are needed for the amplitude describing the process 
$\pi^- K^+ \to \pi^0 K^0$ at one--loop level:
\beqa
{\cal L}^{(4)}_{\rm str} &=& 
    L_3 \,\langle D_\mu U^\dagger D^\mu U D_\nu U^\dagger D^\nu U \rangle
 +  L_4 \,\langle D_\mu U^\dagger D^\mu U \rangle 
      \langle \chi^\dagger U + \chi U^\dagger \rangle \nonumber\\
&+& L_5 \,\langle D_\mu U^\dagger D^\mu U
      \Bigl( \chi^\dagger U + U^\dagger \chi \Bigr) \rangle
 +  L_6 \,\langle \chi^\dagger U + \chi U^\dagger \rangle^2 \nonumber\\
&+& L_7 \,\langle \chi^\dagger U - \chi U^\dagger \rangle^2
 +  L_8 \,\langle \chi^\dagger U \chi^\dagger U + \chi U^\dagger \chi U^\dagger \rangle ~.
\label{L4}
\eeqa
With regard to the analytic formulae given for the S--wave scattering length
in app.~\ref{app:Swave}, we remark that only three of the LECs defined in
eq.~(\ref{L4}) play a role for this quantity at leading order in isospin violation.
Terms depending on the Zweig rule suppressed constants $L_4$ and $L_6$ as well as on $L_7$
are of higher order in isospin breaking.
In addition, the following electromagnetic counterterms given in \cite{Urech}
are needed:
\beqa
{\cal L}^{(4)}_{\rm em} &=& 
    K_1\, F^2 \langle D_\mu U^\dagger D^\mu U \rangle \langle Q^2 \rangle
 +  K_2\, F^2 \langle D_\mu U^\dagger D^\mu U \rangle \langle QUQU^\dagger \rangle
\nonumber\\
&+& K_3\, F^2 \Bigl( \langle D_\mu U^\dagger QU \rangle 
                     \langle D^\mu U^\dagger QU \rangle
                 +   \langle D_\mu UQU^\dagger \rangle 
                     \langle D^\mu UQU^\dagger \rangle \Bigr)
\nonumber\\
&+& K_4\, F^2 \langle D_\mu U^\dagger QU \rangle \langle D^\mu UQU^\dagger \rangle
 +  K_5\, F^2 \langle \Bigl( D_\mu U^\dagger D^\mu U + D_\mu U D^\mu U^\dagger \Bigr)
                         Q^2 \rangle 
\nonumber\\
&+& K_6\, F^2 \langle D_\mu U^\dagger D^\mu UQU^\dagger QU 
                  + D_\mu U D^\mu U^\dagger QUQU^\dagger \rangle
 +  K_7\, F^2 \langle \chi^\dagger U + \chi U^\dagger \rangle \langle Q^2 \rangle
\nonumber\\
&+& K_8\, F^2 \langle \chi^\dagger U + \chi U^\dagger \rangle 
              \langle QUQU^\dagger \rangle
 +  K_9\, F^2 \langle \Bigl( \chi^\dagger U + U^\dagger \chi + \chi U^\dagger + U \chi^\dagger  \Bigr) 
                         Q^2 \rangle
\nonumber\\
&+& K_{10}\, F^2 \langle \Bigl( \chi^\dagger U + U^\dagger \chi \Bigr) QU^\dagger QU 
                       + \Bigl( \chi U^\dagger + U \chi^\dagger \Bigr) QUQU^\dagger \rangle
\nonumber\\
&+& K_{11}\, F^2 \langle \Bigl( \chi^\dagger U - U^\dagger \chi \Bigr) QU^\dagger QU 
                       + \Bigl( \chi U^\dagger - U \chi^\dagger \Bigr) QUQU^\dagger \rangle
\nonumber\\
&+& K_{15}\, F^4 \langle QUQU^\dagger \rangle ^2 
 +  K_{16}\, F^4 \langle QUQU^\dagger \rangle \langle Q^2 \rangle   ~.
\label{L4em}
\eeqa
Also the dependence on a few of these ($K_{7/8}$, $K_{15/16}$) cancels in the
S--wave scattering length.
One more term (proportional to $K_{12}$ in \cite{Urech}) 
that is needed in principle for the renormalization of the decay 
constants of charged mesons can be omitted
in case one uses physical values for decay constants
with electromagnetic effects already subtracted (see section~\ref{sec:S4}).

\medskip \noindent
For numerical evaluation, we use the central values and error estimates for the
hadronic low--energy constants as given in \cite{daphne}. For the
electromagnetic ones, we use the estimates obtained via resonance saturation
in \cite{BauUre}, and add error bars of natural size ($\pm 1/16\pi^2$)
uniformly.


\section{Isospin violation at tree level}

\noindent
Pion--kaon scattering in the isospin limit can be described by two independent
amplitudes $T^{1/2}$ and $T^{3/2}$, corresponding to isospin 1/2 and 3/2, respectively. 
It is sometimes convenient to combine these into isospin--even and --odd amplitudes
$T^\pm$ which are defined by
\beq
T_{\alpha\beta} = \delta_{\alpha\beta} T^+ +\frac{1}{2} [\tau_\alpha,\tau_\beta] T^-  
\eeq
($\alpha$, $\beta$ refer to the isospin indices of the pions)
and which can be related to the amplitudes of definite isospin via
\beqa
3 T^+ &=& T^{1/2} + 2T^{3/2} ~, \\
3 T^- &=& T^{1/2} - T^{3/2} ~.
\eeqa
All these amplitudes can be expressed in terms of the usual Mandelstam variables
$s$, $t$, $u$. They can be decomposed into partial waves $t_l^I(s)$ using
\beq
T^I(s,t)=16\pi \sum_l (2l+1) t_l^I(s) P_l(z) ~,
\eeq
where $z=\cos\theta$ denotes the scattering angle in the center--of--mass system,
and $P_l(z)$ are the Legendre polynomials.
Close to threshold, one can parameterize the real parts of the partial wave amplitudes in terms
of scattering lengths ($a_l$) and effective ranges ($b_l$),
\beq
{\rm Re}\,t_l^I(s) = \frac{\sqrt{s}}{2} \, \Bigl( |\qin||\qout| \Bigr)^l 
   \left\{ a_l^I + b_l^I |\qin|^2 + {\cal O} \left(|\qin|^4\right) \right\} ~,
\label{def:thrpar}
\eeq
where we have already accounted for the possibility of different masses of the incoming
and outgoing particles and therefore for different incoming and outgoing center--of--mass
momenta,
\beqa
|\qin | &=& \frac{\sqrt{(s-(M_K^{\rm in } \!-\! M_\pi^{\rm in })^2)
                        (s-(M_K^{\rm in } \!+\! M_\pi^{\rm in })^2)}}{2\sqrt{s}} ~, \label{qin}
\\
|\qout| &=& \frac{\sqrt{(s-(M_K^{\rm out} \!-\! M_\pi^{\rm out})^2)
                        (s-(M_K^{\rm out} \!+\! M_\pi^{\rm out})^2)}}{2\sqrt{s}} ~, \label{qout}
\eeqa
where $M_{\pi/K}^{\rm in/out}$ refer to the physical masses of the incoming and outgoing
pions and kaons, respectively. In the isospin limit, eq.~(\ref{def:thrpar}) collapses to the usual definition.
We note that the real part of a total amplitude at threshold 
(i.e.\ for $|\qin|=0$) is linked to the S--wave scattering length by
\beq
{\rm Re}\,T^I_{\rm thr} = 8\pi \Bigl(M_K^{\rm in} \!+\! M_\pi^{\rm in}\Bigr)\, a_0^I 
 + {\cal O}\left(|\qin|^2,|\qin||\qout|\right) ~. \label{threxp}
\eeq

\medskip \noindent
At tree level, the isospin--even and --odd pion--kaon scattering lengths
are given by \cite{We66,Griffith}
\beq
a_0^+ = 0 ~, \quad 
a_0^- = \frac{M_\pi M_K}{8\pi F_\pi^2(M_\pi+M_K)} = 70.8\times 10^{-3}/M_\pi ~.
\eeq

\medskip \noindent
The different physical pion--kaon channels receive corrections to 
their isospin symmetric scattering lengths when taking into account the mass
differences as well as insertions stemming from the electromagnetic term
in eq.~(\ref{L2}). We collect the isospin breaking parameters
as $\delta \in \{ m_u\!-\!m_d,e^2 \}$ and expand the corrections to
the scattering lengths up to order $\delta$.
The following conventions were used:
\begin{itemize}
\item The isospin symmetry limit is defined according to e.g.\ \cite{GGLR}, 
i.e.\ we express everything in terms of the \emph{charged} meson masses 
$\Mp$, $\MK$.
Note that these are not exactly the natural choices from the point of view of a
chiral analysis of the meson masses, in which case one would prefer to take
the neutral pion mass as a reference (which is indeed done in \cite{MMS}),
as one has $\Mn^2=2B\hat{m}$ (with $\hat{m}=\frac{1}{2}(m_u+m_d)$) 
to good accuracy, neglecting only tiny corrections
of order $(m_u-m_d)^2$. However, one would consequently have to resolve to
using a ``non--physical'' kaon mass $M_K^2=B(\hat{m}+m_s)\approx 495$~MeV.
Arguments \emph{for} the use of the charged pion and kaon masses are the 
correct kinematics for the experimentally accessible channels (which use incoming
charged particles), and the fact that the existing study of pion--kaon scattering
in ChPT (in the isospin symmetry limit) \cite{BKM} also employs these choices
for numerical evaluation.
\item As finally we want to evaluate isospin violating effects up to order
$e^2$ and $m_u\!-\!m_d$, we can neglect the strong pion mass difference 
(of order $(m_u\!-\!m_d)^2$). Quark mass insertions can then, at leading
order, be expressed by physical meson masses according to
\beqa
\hat{m}  &\to& \frac{\Mn^2}{2B} ~, \quad
m_d-m_u \to \frac{\MKn^2 \!-\! \MK^2 \!+\! \Mp^2 \!-\! \Mn^2}{B} ~, \nonumber\\
m_s &\to& \frac{\MK^2 \!+\! \MKn^2 \!-\! \Mp^2}{2B} ~, \quad
Z \to \frac{\Mp^2 \!-\! \Mn^2}{2e^2F_\pi^2} ~.
\label{quarkMeson}
\eeqa
\item One important isospin breaking effect due to the light quark mass difference
is $\pi^0 \eta$--mixing: the $\pi^0$ and $\eta$ fields are given in terms of
the SU(3) eigenstates $\phi_3$, $\phi_8$ according to
\beq
\biggl(\! \begin{array}{c} \pi^0 \\ \eta \end{array} \!\biggr)
= \biggl(\! \begin{array}{cc} \cos\epsilon & \sin\epsilon \\ 
                             -\sin\epsilon & \cos\epsilon \end{array} \!\biggr)
  \biggl(\! \begin{array}{c} \phi_3 \\ \phi_8 \end{array} \!\biggr) ~,
\label{pi0etamix}
\eeq
where the $\pi^0 \eta$ mixing angle $\epsilon$ is given by
\beq
\epsilon = \frac{1}{2} \arctan \left(\frac{\sqrt{3}}{2}\frac{m_d-m_u}{m_s-\hat{m}}\right) ~.
\eeq
Replacing the quark masses by meson masses according to eq.~(\ref{quarkMeson}),
one finds the numerical value $\epsilon = 1.00 \times 10^{-2}$.
We find it convenient to express all corrections due to the light quark mass difference
in terms of this mixing angle $\epsilon$, which is of course of order $m_u\!-\!m_d$.
\end{itemize}
Note finally that we only display scattering lengths for processes involving
kaons of positive strangeness ($K^+$, $K^0$) as the strong and electromagnetic
interactions obey charge conjugation invariance, such that scattering lengths for all 
channels involving scattering of $K^-$ or $\bar{K}^0$ can be obtained from those given below.
We find:
\beqa
a_0 \Bigl( \pi^+ K^+ \to \pi^+ K^+ \Bigr) &=& 
  a_0^{3/2} \left\{ 1 - \frac{2ZF_\pi^2}{M_\pi M_K}\,e^2 \right\} 
  + {\cal O}(\delta^2)  \nonumber \\
&=& a_0^{3/2} \Bigl\{ 1 - 0.018  \Bigr\} 
  + {\cal O}(\delta^2)  ~, \label{a++++} \\
a_0 \Bigl( \pi^+ K^0 \to \pi^+ K^0 \Bigr) &=& 
  \Bigl(a_0^+ + a_0^- \Bigr) \left\{ 1 
  + \frac{2(M_K \!-\! M_\pi)M_\pi}{\sqrt{3}M_K^2} \, \epsilon
  - \frac{Z\,M_\pi F_\pi^2}{M_K^2(M_K \!+\! M_\pi)} \,e^2 \right\} 
  + {\cal O}(\delta^2)  \nonumber \\
&=& \Bigl(a_0^+ + a_0^- \Bigr) \Bigl\{ 1 
  + 0.002 - 0.001 \Bigr\} 
  + {\cal O}(\delta^2) ~,  \\
a_0 \Bigl( \pi^+ K^0  \to \pi^0 K^+ \Bigr) &=&
  \sqrt{2} \, a_0^- \left\{1  
  - \frac{M_K^2 \!-\! 2M_KM_\pi \!+\! 2M_\pi^2}{\sqrt{3}M_K^2} \,\epsilon
  - \frac{Z\,(M_K^2 \!+\! M_\pi^2) F_\pi^2}{M_K^2M_\pi(M_K \!+\! M_\pi)} \,e^2  \right\} 
\nonumber \\ && \qquad \qquad \qquad
  + {\cal O}(\delta^2) \nonumber \\
&=&  \sqrt{2} \, a_0^- \Bigl\{1  
  - 0.003 - 0.008 \Bigr\} + {\cal O}(\delta^2) ~, \\
a_0 \Bigl( \pi^0 K^+  \to \pi^0 K^+ \Bigr) &=& 
  a_0^+ - \frac{M_K^2}{4\sqrt{3}\pi F_\pi^2 (M_K \!+\! M_\pi)} \,\epsilon
  + {\cal O}(\delta^2)  \nonumber\\
&=&  a_0^+ - 2.9 \times 10^{-3}/M_\pi  
  + {\cal O}(\delta^2) ~, \label{a0+0+} \\
a_0 \Bigl( \pi^- K^+  \to \pi^- K^+ \Bigr) &=& 
  \Bigl( a_0^+ + a_0^- \Bigr) \left\{1 + \frac{2Z F_\pi^2}{M_\pi M_K} \, e^2 \right\} 
  + {\cal O}(\delta^2)  \nonumber\\
&=& \Bigl( a_0^+ + a_0^- \Bigr) \Bigl\{1 + 0.018 \Bigr\} 
  + {\cal O}(\delta^2) ~, \label{a-+-+} \\
a_0 \Bigl( \pi^- K^+  \to \pi^0 K^0 \Bigr) &=& 
  -\sqrt{2}\, a_0^- \left\{1 +\frac{\epsilon}{\sqrt{3}}
  + \frac{Z F_\pi^2}{M_\pi M_K} \, e^2 \right\} 
  + {\cal O}(\delta^2) \nonumber\\
&=&  -\sqrt{2}\, a_0^- \Bigl\{1 + 0.006 + 0.009 \Bigr\} 
  + {\cal O}(\delta^2) ~, \label{a-+00}\\
a_0 \Bigl( \pi^0 K^0  \to \pi^0 K^0 \Bigr) &=& 
  a_0^+ + \frac{M_K^2}{4\sqrt{3}\pi F_\pi^2 (M_K \!+\! M_\pi)}\, \epsilon
  + {\cal O}(\delta^2)  \nonumber\\
&=&   a_0^+ + 2.9 \times 10^{-3}/M_\pi + {\cal O}(\delta^2) ~, \label{a0000} \\
a_0 \Bigl( \pi^- K^0  \to \pi^- K^0 \Bigr) &=& 
  a_0^{3/2} \left\{1 + \frac{2(M_K \!-\! M_\pi)M_\pi}{\sqrt{3}M_K^2} \,\epsilon
  - \frac{Z M_\pi F_\pi^2}{M_K^2(M_K \!+\! M_\pi)} \, e^2 \right\} 
  + {\cal O}(\delta^2) \nonumber \\
&=& a_0^{3/2} \Bigl\{1 + 0.002 - 0.001 \Bigr\} + {\cal O}(\delta^2) ~. 
\eeqa
For the elastic charged channels, eqs.~(\ref{a++++}), (\ref{a-+-+}),
subtraction of the one--photon exchange Born term is required in order
to arrive at the above results (see~\cite{KU} for the comparable $\pi\pi$ case).
These charged--particle Coulomb interactions in principle require a separate treatment
tailored to scattering or bound--state problems. 
(For the $\pi\pi$ scattering problem, this is discussed e.g.\ in~\cite{RS}.)
We remark that isospin violation effects can only be given in absolute size
for those amplitudes which are (in the isospin limit) proportional to $T^+$, 
as $a_0^+$ vanishes at tree level. The one--loop corrections to $a_0^+$ 
are rather large, $a_0^{+(4)}=31.9 \times 10^{-3}/M_\pi$ \cite{BKM}, 
but still then the corrections displayed in eqs.~(\ref{a0+0+}), (\ref{a0000})
amount to 10\% effects.  
(The caveat is, of course, that isospin violating contributions 
at one--loop level might reduce the leading--order terms considerably.)
In fact, this is neatly comparable to pion--nucleon scattering: as pointed out by
Weinberg, the isoscalar $\pi N$ amplitude vanishes at leading chiral order
\cite{We66} and is therefore prone to display large isospin violation effects
\cite{We77}, which have indeed been confirmed in \cite{piN-IV}.

\medskip \noindent
It is instructive to compare the results above also to the corresponding corrections
calculated in \cite{KU} for $\pi\pi$ scattering. First of all, $m_u-m_d$ effects
are suppressed in SU(2), such that the only isospin breaking corrections of interest
in the $\pi\pi$ case are of electromagnetic origin. In contrast, strong isospin breaking
appears at leading order in $\pi K$ scattering and, as shown above, is potentially of
equal significance as the electromagnetic effects. On the other hand, the authors
of \cite{KU} find corrections at tree level of the order of 6\%, while the largest
corrections given above are 1.8\% (for the elastic charged meson channels). 
This is not due to a particular ``smallness'' of e.g.\ the electromagnetic contribution
for $\pi K$ scattering, but due to the fact that the (isospin conserving) S--wave scattering
length is larger, $a_0(\pi K) \sim M_\pi M_K$ in contrast to $a_0(\pi\pi) \sim M_\pi^2$.
As already hinted at in the introduction, processes with a conserved number of strange quarks
(such as $\pi K$ scattering) can also be studied in a chiral SU(2) theory in which
Green's functions are only expanded around the limit $m_u=m_d=0$, while $m_s$ is held fixed
at its physical value. Consequently, only the pion is considered to be ``light'', while
the kaon is now a heavy particle and has to be treated in analogy to e.g.\ nucleons in ChPT,
resulting in what one might call ``heavy--kaon'' ChPT \cite{Roessl,Ou}. 
The $\pi K$ Lagrangian then starts out at order $q$, as does the (isovector) $\pi K$
scattering length (see again the analogy to $\pi N$ scattering). 
In such a counting scheme, electromagnetic corrections to $\pi K$ scattering lengths
are therefore necessarily suppressed by one chiral order (as they scale like $e^2$), 
which sheds some light on why these corrections are smaller here than for $\pi\pi$ scattering.

\medskip \noindent
We want to emphasize that it is in general impossible to arrive at the results above
by assuming the dominance of certain ``effects'' for the violation of isospin symmetry,
e.g.\ studying the implications of meson mass splittings or $\pi^0 \eta$ mixing alone.
From the point of view of ChPT, such a partial analysis is not very useful and can lead
to strongly misleading results. In order to demonstrate this, we show such a decomposition
(artificial though this really is) for the specific channel studied in more detail
in the following chapter, $\pi^- K^+ \to \pi^0 K^0$. We distinguish ``kinematical'' effects
(due to meson mass splittings, entering essentially 
via corrections to the values of $u$ and $t$ at threshold), 
``$\pi^0 \eta$ mixing'' effects which modify the isospin symmetric amplitude by
factors of $\sin\epsilon$ or $\cos\epsilon$,
``quark mass insertions'' for the four--meson vertex
(which vanish in the isospin limit for the channel in question),
and ``electromagnetic (em) insertions'' ($\sim Z$) for the four--meson vertex.
The relative corrections can then be decomposed as follows:
\beqa
{\rm strong:} && \quad\quad\!
\frac{\epsilon}{\sqrt{3}} \,\,=\,\,
\frac{\epsilon}{\sqrt{3}} \,\Biggl\{ 
\underbrace{1 -\frac{M_K}{M_\pi}}_{\rm kinematical}
\,+\,\underbrace{\frac{2M_K^2+M_\pi^2}{3M_KM_\pi}}_{\pi^0\eta-{\rm mixing}}
\,+\,\underbrace{\frac{M_K^2-M_\pi^2}{3M_KM_\pi}}_{\rm quark\,\,mass} \Biggr\} ~,\quad
\\
{\rm electromagnetic:} &&  \frac{Ze^2F_\pi^2}{M_KM_\pi} \,\,=\,\,
\underbrace{\frac{Ze^2F_\pi^2}{2M_KM_\pi}}_{\rm kinematical}
\,+\, \underbrace{\frac{Ze^2F_\pi^2}{2M_KM_\pi}}_{\rm em\,\,insertions} ~.
\eeqa
It is obvious from the above that for the strong isospin violating contributions,
individual ``effects'' are much larger than the total sum.


\section{$\pi^- K^+ \to \pi^0 K^0$ at one--loop level}
\subsection{S--wave scattering length up to fourth order\label{sec:S4}}

\begin{figure}[htb]
\vskip 0.5cm
\centerline{
\epsfysize=3.5cm
\epsffile{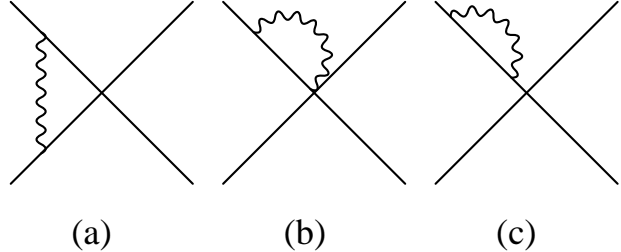}
}
\caption{Virtual photon loop diagrams for $\pi^- K^+ \to \pi^0 K^0$. 
Crossed diagrams are not shown. Diagram (c) denotes effects entering
the scattering amplitude via wave function renormalization of the incoming
charged mesons.
\label{fig:photonloops}
}
\end{figure}

\noindent 
In this section, we work out the complete one--loop corrections including
isospin breaking for the pion--kaon channel that is relevant for the lifetime
of $\pi K$ atoms, $\pi^- K^+ \to \pi^0 K^0$. At this level, virtual photon loops
play a role and enter the amplitude both via wave function renormalization and
electromagnetic corrections to the (lowest order) strong vertex,
see fig.~\ref{fig:photonloops}.
As is well known, the photon exchange diagram (a) in fig.~\ref{fig:photonloops} 
induces a pole at threshold, the so-called Coulomb pole, 
which resides in the loop function $G^{\pi K\gamma}(s)$
discussed in detail in app.~\ref{app:GpKg}. 
This Coulomb pole, however, can be calculated and subtracted from the amplitude unambiguously.
The threshold expansion eq.~(\ref{threxp}) therefore has to be modified
for $\pi^- K^+ \to \pi^0 K^0$ in the presence of virtual photons
in order to sensibly define corrections to the S--wave scattering length:
\beq
{\rm Re}\,T 
= \frac{e^2 \pi \MK \Mp \,a_0^{(2)}}{|\qin|}
+8\pi \Bigl(\MK \!+\! \Mp\Bigr)\, a_0 + {\cal O}(|\qin|) ~,
\label{Cpole}
\eeq
where $a_0^{(2)}$ denotes the S--wave scattering length in its tree level approximation
(given explicitly up to order $\delta$ in eq.~(\ref{a-+00})), and $a_0$ is 
the S--wave scattering length to be defined here, including all corrections.

\medskip \noindent
A further problem occurs when including virtual photon loops: the amplitude becomes
infrared divergent. We will give details on how to deal with these infrared divergences
in the following section. For the moment we only note that these vanish
at threshold and therefore do not affect the definition of the S--wave scattering 
length as given in eq.~(\ref{Cpole}).

\medskip \noindent
Analytic formulae for the S--wave scattering length at one--loop accuracy
are given in app.~\ref{app:Swave}.
In order to arrive at these, the following effects were taken into account, 
as well as the following conventions were used in addition to those already
mentioned in the previous section:
\begin{itemize}
\item In order to have a strong check on the scattering amplitude, we have
checked the cancellation of divergences in the \emph{exact}
expressions, i.e.\ without any expansion in $\epsilon$ and $e^2$. 
For this purpose, we have used the $\beta$--functions for the various
counterterms as given in~\cite{NPB250,Urech}. 
\item
For quark mass insertions at tree level replaced by meson masses according to
eq.~(\ref{quarkMeson}), the renormalization
of these meson masses has to be taken into account properly.
\item The mass of the $\eta$ is always taken as given by the Gell-Mann--Okubo
relation, which reads
\beq
3M_\eta^2 = 2\MK^2+2\MKn^2-2\Mp^2+\Mn^2 
\eeq
when taking into account isospin violation up to order $e^2$ and $m_u\!-\!m_d$.
Deviations from this relation are beyond the accuracy considered in this paper,
as the $\eta$--mass only enters via loop diagrams.
\item 
\begin{figure}[htb]
\centerline{
\epsfysize=4.cm
\epsffile{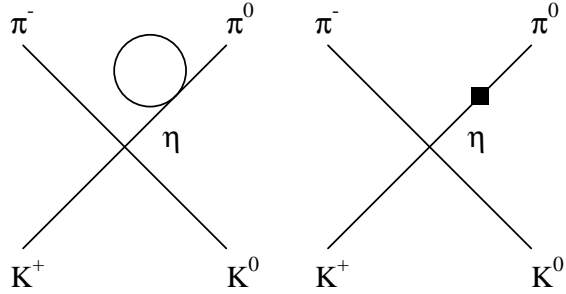}
}
\caption{Diagrams contributing to $\pi^0 \eta$ mixing at next--to--leading order.
The labels ``$\pi^0$'' and ``$\eta$'' refer to the tree--level mass eigenstates.
The black square denotes a fourth--order insertion (strong or electromagnetic).
\label{fig:pi0eta}
}
\end{figure}
$\pi^0 \eta$--mixing has to be taken into account at next--to--leading order. 
Details on how this is to be done can be found 
e.g.\ in~\cite{eta3pi,NeuRup,ABT,EckMul}.
It is probably easiest to calculate the diagrams in fig.~\ref{fig:pi0eta}
directly in order to account for mixing beyond leading order.
We have done the calculations using tree--level mass eigenfields for $\pi^0$ and $\eta$ throughout.
The off--diagonal element of the self--energy matrix (in the basis of tree--level
eigenfields) may be written as 
$\Sigma_{\pi^0\eta}(q^2) = Z_{\pi^0\eta} q^2 + Y_{\pi^0\eta}$,
such that $Z_{\pi^0\eta}=\partial \Sigma_{\pi^0\eta} / \partial q^2$
can be thought of as the off--diagonal analogy to the field renormalization factors.
Denoting the tree--level amplitude for $\pi^- K^+ \to \eta K^0$ by $T_\eta$,
we can write the additional mixing diagrams as
\beqa
&&T_\eta \times \frac{1}{\Mn^2 \!-\! M_\eta^2} \times \Sigma_{\pi^0\eta} \left(q^2=\Mn^2\right)
\nonumber\\
&=& T_\eta \times \Biggl\{ \frac{Z_{\pi^0\eta}}{2} 
+ \underbrace{\frac{1}{\Mn^2 \!-\! M_\eta^2} \times 
  \Sigma_{\pi^0\eta} \left(q^2=\frac{1}{2}\left(\Mn^2\!+\!M_\eta^2\right) \right)}
  _{\epsilon^{(4)}}
\,\Biggr\} ~. \label{loopmix}
\eeqa
$\epsilon^{(4)}$ thus defined is the correction to the tree level $\pi^0\eta$ mixing angle 
up to order ${\cal O}(q^4)$, as given e.g.\ in \cite{EckMul}.
In contrast to the wave function renormalization factor $Z_{\pi^0\eta}$, $\epsilon^{(4)}$
is finite.

\item When expressing the low--energy constant $F$ in terms of physically
observable decay constants, we always renormalize it such as to obtain $F_\pi$
(as opposed to $F_K$, say). On tree level, the difference between using a 
normalization of the amplitude equal to $1/F_\pi^2$ and $1/F_\pi F_K$
amounts to nothing more but shifting certain contributions from tree 
to one--loop level, which makes no difference in the sum. However, it turns
out that the one--loop corrections e.g.\ for the scattering lengths
are much smaller when normalizing the tree level expression to $1/F_\pi^2$,
which means that in the seemingly more ``symmetric'' case of using $1/F_\pi F_K$,
the bulk of the (large) corrections at ${\cal O}(q^4)$ have nothing to do
with corrections to the scattering dynamics, but only with the renormalization
of $F_K$. 

\medskip \noindent
For the one--loop contributions, the difference between $F_\pi$ and $F_K$ is
of higher order, whereas the numerical difference is potentially significant
(noting that these scale as $1/F^4$). Regarding $F_0 \simeq 88$~MeV~\cite{Annals}
as the ``natural'' constant when writing down a scattering amplitude,
$F_\pi$ seems more appropriate than $F_K$.

\medskip \noindent
Finally heavy--kaon ChPT, which we alluded to before already, also 
provides some insight on $F_\pi$ versus $F_K$: there is of course no kaon decay
constant in a kaon--number conserving theory, hence $F_\pi$ is the only meson decay
constant available. As heavy--kaon ChPT is supposed to have a better convergence
behavior than SU(3) ChPT, but ultimately is equivalent up to matching of the
different sets of LECs, it seems reasonable also from this perspective
to only use $F_\pi$ in the $\pi K$ amplitudes.
\item We use the charged pion decay constant in the absence of electromagnetism,
as the value commonly used for the physical pion decay constant, 
$F_\pi=92.4$~MeV~\cite{Holstein}, 
was extracted from \emph{charged} pion decays
with electromagnetic corrections already taken care of. 
The relation to the bare decay constant $F$ is therefore as given in~\cite{NPB250},
and independent of photon loop effects or electromagnetic counterterms.
\end{itemize}
We display the isospin conserving and violating contributions, at tree and
one--loop level, relative to the isospin symmetric tree level result (which
we just denote by $a_0^{(2)}$ here for reasons of brevity).
Numerically, these add up as follows: 
\beqa
a_0^{(2)} &=& -100.1 \times {10^{-3}}/{\Mp} ~, \nonumber\\
a_0  &=&  a_0^{(2)} \,\Bigl\{ \, 1 
         \,+\, \underbrace{0.006}_{{\cal O}(\epsilon)}      
         \,+\, \underbrace{0.009 }_{{\cal O}(e^2)} \nonumber\\
&&   \quad +\, \underbrace{(0.121 \pm 0.009)}_{{\cal O}(p^4)} 
         \,+\, \underbrace{(0.008 \pm 0.002)}_{{\cal O}(p^2 \epsilon)}    
         \,-\, \underbrace{(0.009 \pm 0.008)}_{{\cal O}(p^2 e^2)} \, \Bigr\} ~.
\label{a0num}
\eeqa
All errors quoted for the different contributions above
are due to uncertainties in the respective
strong and electromagnetic LECs as given in \cite{daphne,BauUre}.
The following comments are in order:
\begin{itemize}
\item The isospin symmetric corrections at one--loop order are
moderate, given the expected slow convergence of chiral SU(3). 
As detailed above, this
depends heavily on our choice to normalize the tree amplitude
by $1/F_\pi^2$ and not $1/F_\pi F_K$. 
In fact, expanding the one--loop contributions in powers of $M_\pi$ 
(equivalently to the heavy--kaon approach), one finds that
they are suppressed essentially by a factor of $M_\pi^2$ 
(and not by $M_\pi M_K$ or $M_K^2$) compared to the tree level values,
which is what one would expect from an SU(2) expansion and which
accounts for the smallness of the one--loop corrections.
\item 
Although they are very small corrections to the scattering length, 
one might worry about the fact that tree and one--loop level
isospin breaking corrections (strong and electromagnetic effects taken separately)
are of equal size. 
In order to understand the relative largeness of these effects, 
it is instructive to expand the terms
given in app.~\ref{app:Swave} in powers of $M_\pi$. 
Isospin violating terms are suppressed
in comparison to isospin conserving ones at the same chiral order
by factors which happen to be small, but are of chiral order 1. 
On tree level, these suppression factors for $\pi^- K^+ \to \pi^0 K^0$ 
are $\epsilon$ and $e^2/M_\pi M_K$, respectively, see eq.~(\ref{a-+00}).
On one--loop level, however, where the isospin symmetric loop corrections
are particularly small as discussed above, the isospin violating ones 
are suppressed with respect to these only by factors of $\epsilon\, M_K/M_\pi$
and $e^2\, M_K/M_\pi^3$ (at leading order in $M_\pi$). 

\medskip \noindent
On the other hand, combining strong and electromagnetic isospin breaking effects,
the corrections at one--loop level cancel to a large extent, leaving mainly an uncertainty
due to a lack of sufficiently precise knowledge of the relevant counterterms.
This cancellation does not take place at tree level.
We have checked numerically that no significant
errors arise from truncating the ``exact'' expressions at leading orders
in the isospin breaking parameters $\epsilon,~e^2$. In fact, the corrections
arising thereof are at least one order of magnitude smaller than the accuracy displayed
in eq.~(\ref{a0num}). 
\item With the large number of electromagnetic counterterms, one might have
suspected that one cannot make any sensible prediction about electromagnetic
effects at all. While the uncertainty quoted above is as large as the absolute
size of the electromagnetic correction at one--loop level, it is, on the other hand, still not
larger than the uncertainty stemming from our lack of knowledge of the values
for the hadronic low--energy constants.
\end{itemize}


\subsection{Infrared divergences and $\pi^- K^+ \to \pi^0 K^0 \gamma$}

\noindent
As mentioned in the previous section, photon loop corrections induce
infrared divergences in the scattering amplitude.
It is well known how to handle this problem, namely by introducing
a small photon mass $m_\gamma$ which brings the infrared divergences into a form
$\sim \log m_\gamma$. 
Such terms enter both via wave function renormalization of charged mesons
and via the loop function $G^{\pi K \gamma}(s)$, as discussed in app.~\ref{app:GpKg}.
One then has to include the corresponding radiative process 
($\pi^- K^+ \to \pi^0 K^0 \gamma$ for the case in question) up to some maximal
photon energy, given either by the maximal energy available due to kinematics, 
or by some experimental detector resolution $\Delta E$ below which soft photon
radiation cannot be discriminated from the non--radiative process. 
We are interested in the threshold region, where the kinematical limit for the
photon energy,
\beq
E_\gamma^{\rm max} = \frac{s-(\MKn \!+\! \Mn)^2}{2\sqrt{s}} ~,
\eeq
numerically amounts to 0.6~MeV, such that it seems reasonable to integrate the full
cross section without making additional cuts on the photon energy.
The infrared divergences cancel upon adding up the cross sections for both
non--radiative and radiative processes. This is demonstrated in some detail in
app.~\ref{app:radXsec:div}.

\medskip \noindent
We have emphasized in the previous chapter that the infrared divergences induced
by photon loops vanish at threshold, such that one might even omit all these
considerations when one is exclusively interested in the S--wave scattering length.
The authors of \cite{KU} have argued, however, that one should rather \emph{define}
the S--wave scattering length from an infrared--finite quantity, which would be the 
combined total cross section. 
For an amplitude which can be expanded
according to eq.~(\ref{threxp}), the total cross section at threshold is given
in terms of the S--wave scattering length as
\beq
\sigma_{\rm thr} = \frac{|\qout|}{|\qin|} \, 4\pi (a_0)^2 ~.
\label{Xsec-a0}
\eeq
One may now replace the cross section on the left--hand side of eq.~(\ref{Xsec-a0})
by the combined total cross section $\sigma + \sigma^\gamma$ and define
$a_0$ via eq.~(\ref{Xsec-a0}). 

\medskip \noindent
This is what was suggested in \cite{KU} for the
similar case of the scattering process $\pi^+ \pi^- \to \pi^0 \pi^0 (\gamma)$,
where the authors claim to find a shift in the (redefined) scattering length induced
by the (infrared--finite) remainder of the radiative cross section at threshold.
We have however recalculated the cross section for $\pi^+ \pi^- \to \pi^0 \pi^0 \gamma$
at threshold and find that it vanishes. 
This is in our opinion rather obvious for the isospin symmetric case, where the
neutral pions in the final state are of the same mass as the charged incoming ones. 
One would therefore expect the corrections to the scattering length due to 
soft photon radiation to be small
compared to other isospin breaking effects, as they should be suppressed by at least
two powers in isospin breaking parameters (a factor $e^2$ in the cross section
due to the photon coupling, and at least one power of the pion mass difference). 
The fact that this cross section vanishes exactly at threshold in the $\pi\pi$ case
is due to the enhanced symmetry of this reaction, see app.~\ref{app:radXsec:thr}.

\medskip \noindent
We do not find the cross section for $\pi^- K^+ \to \pi^0 K^0 \gamma$ to vanish
exactly at threshold, but to be highly suppressed: the leading contribution
is of order $\delta^4$ in isospin breaking parameters, which is clearly way
beyond the level of accuracy we achieve in our calculation of isospin breaking
effects in the non--radiative amplitude.
Details for the calculation can be found in app.~\ref{app:radXsec:thr}, the result 
found there is
\beqa
\sigma^\gamma_{\rm thr} &=&
\frac{e^2}{105(2\pi)^3 F_\pi^4} \frac{|\qout|}{|\qin|} \frac{\MK\Mp}{(\MK\!+\!\Mp)^3}
\Bigl( \MK \!-\! \MKn \!+\! \Mp \!-\! \Mn \Bigr)^3 + {\cal O}(\delta^5) ~.
\eeqa
In addition to the high power in $\delta$, there is even an unnatural suppression
in the mass difference $(\MK\!-\!\MKn\!+\!\Mp\!-\!\Mn)$ which is due to a numerical
cancellation of strong and electromagnetic isospin breaking effects.
If one calculates a shift in the scattering length defined via eq.~(\ref{Xsec-a0})
which is due to the inclusion of the radiative cross section, $a_0 \to a_0 + a_0^\gamma$,
the quantity $a_0^\gamma$ is found to be
\beqa
a_0^\gamma &=& -\frac{e^2}{840\sqrt{2}\pi^3F_\pi^2} 
\frac{\bigl( \MK \!-\! \MKn \!+\! \Mp \!-\! \Mn \bigr)^3}{(\MK\!+\!\Mp)^2} +{\cal O}(\delta^5)
\nonumber\\
&=& -2.2 \times 10^{-14} /\Mp 
\,\,=\,\, 2.2 \times 10^{-13} \times  a_0^{(2)}
\quad,
\eeqa
which is a suppression of about eleven orders of magnitude compared to the shifts in the scattering
length calculated before. 
We therefore conclude that bremsstrahlung corrections are totally negligible for
any analysis of $\pi\pi$ or $\pi K$ bound state experiments.


\subsection{S--wave effective range, P--wave scattering length}

\noindent
Although a precise knowledge of the S--wave scattering length is of highest interest
for $\pi K$ atom studies, one can of course also study isospin breaking corrections
to other threshold parameters, particularly to the S--wave effective range $b_0$
and the P--wave scattering length $a_1$.
We will not show results for these in as much detail as for $a_0$, 
but concentrate on the channel for which we have calculated the
one--loop corrections to isospin breaking. 
We remark that, as indicated in eq.~(\ref{Cpole}), photon loops also
induce a term \emph{linear} in $|\qin|$ in the threshold expansion of the 
scattering amplitude, which we disregard here 
(as it can also be calculated and subtracted unambigously).

\medskip \noindent
The isospin symmetric tree level results for these two quantities are given by
\beqa
b_0^{(2)} \,=\, -\sqrt{2}\, b_0^- &=& 
  -\frac{M_K^2 \!+\! M_\pi^2}{8\sqrt{2}\pi F_\pi^2 M_\pi M_K (M_K \!+\! M_\pi)} ~~, \\
a_1^{(2)} \,=\, -\sqrt{2}\, a_1^- &=& 
  -\frac{1}{24\sqrt{2}\pi F_\pi^2 (M_K \!+\! M_\pi)} ~~.
\eeqa
The isospin violating effects at tree level can then be expressed as
\beqa
b_0 \Bigl( \pi^- K^+ \to \pi^0 K^0 \Bigr) &=& -\sqrt{2}\, b_0^- \, \biggl\{ 1
+ \frac{\epsilon}{\sqrt{3}} 
- \frac{Ze^2F_\pi^2}{M_K^2 \!+\! M_\pi^2} \biggr\}
+ {\cal O}(\delta^2)   ~~, \\
a_1 \Bigl( \pi^- K^+ \to \pi^0 K^0 \Bigr) &=& -\sqrt{2}\, a_1^- \, \Bigl\{ 1 
- \sqrt{3} \,\epsilon \Bigr\} + {\cal O}(\delta^2)  ~~.
\eeqa
No electromagnetic effects affect the P--wave scattering length at this order.

\medskip \noindent
We evaluate the fourth--order contributions to these threshold parameters
numerically only,
therefore we do not differentiate between strong and electromagnetic 
isospin violation at this level.
Again, all corrections are given relative to the isospin symmetric
tree--level result:
\beqa
b_0^{(2)} &=& -54.0 \times {10^{-3}}/{\Mp^3} ~, \nonumber\\
b_0  &=&  b_0^{(2)} \,\Bigl\{ \, 1
         \,+\, \underbrace{0.006}_{{\cal O}(\epsilon)}      
         \,-\, \underbrace{0.002}_{{\cal O}(e^2)} 
         \,+\, \underbrace{(0.012 \pm 0.038)}_{{\cal O}(p^4)} 
         \,-\, \underbrace{(0.013 \pm 0.005)}_{{\cal O}(p^2\delta)} \,\Bigr\} ~,
\\
a_1^{(2)} &=& -4.7 \times {10^{-3}}/{\Mp^3} ~, \nonumber\\
a_1  &=&  a_1^{(2)} \,\Bigl\{ \, 1
         \,-\, \underbrace{0.02}_{{\cal O}(\epsilon)}      
         \,+\, \underbrace{(0.53 \pm 0.14)}_{{\cal O}(p^4)} 
         \,-\, \underbrace{(0.03 \pm 0.01)}_{{\cal O}(p^2\delta)} \,\Bigr\} ~.
\eeqa
We note that the isospin symmetric one--loop corrections to the P--wave scattering length
are rather large compared to the leading term. 
The isospin breaking corrections for both quantities
are of similar size as for $a_0$, only slightly enhanced for $a_1$. 
The main difference is that the cancellation between electromagnetic and $m_u\!-\!m_d$ effects at fourth order
observed for $a_0$ does not occur here, such that isospin breaking at fourth chiral
order is equally important as on tree level.
If we compare the one--loop contributions only, the isospin breaking effects on the 
S--wave effective range appear to be very large, equally large even as the isospin symmetric
one--loop corrections. However, as already hinted at by the error range for the latter, 
this does not result from an unnatural enhancement of isospin violation effects, but from
an accidental suppression of the isospin conserving part. Individual contributions to the
latter (from separate loop diagrams, counterterms) are much larger.

\medskip \noindent
In contrast to both the S--wave and the P--wave scattering lengths, the
S--wave effective range as defined naively from the scattering amplitude
alone is infrared divergent. 
The above results were achieved by setting $m_\gamma$ equal to 
the maximal energy of a bremsstrahlung photon, which at threshold
is roughly $0.6$~MeV. This is what effectively happens when rendering infrared divergent
expressions finite by including the appropriate radiative process, while we 
neglect here any finite contributions from the latter, see the discussion
in the previous section. We also note that a 
redefinition of these higher--order threshold parameters 
from the infrared--finite total cross section 
(as done in the previous section for the S--wave scattering length)
cannot be done unambiguously.


\section{Summary}

\noindent
In this work, we have considered isospin violation in pion--kaon
scattering near threshold. To systematically account for such
effects due to the light quark mass difference as well as
electromagnetic interactions, we have made use of SU(3) chiral perturbation
theory in the presence of virtual photons. The pertinent results of
this investigation can be summarized as follows:
\begin{enumerate}
\item[(1)] Already at tree level, one has strong as well as
electromagnetic isospin violation. We have considered all physical
channels and found that these effects are in general small (at most two percent). 
In particular, the relative corrections are smaller than in the comparable
case of $\pi\pi$ scattering where isospin violation at tree level is
a purely electromagnetic effect. We have also stressed that
considering one particular source of isospin violation only can ensue
very misleading results.
\item[(2)]Because of its relevance to the lifetime of $\pi K$ atoms,
we have considered the one--loop corrections to the S--wave scattering
length for the process $\pi^- K^+ \to \pi^0 K^0$. The fourth--order
isospin symmetric corrections are moderate and the isospin violation
effects at this order cancel to a large extent (for the central values of the
low--energy constants used here). 
The uncertainty in these isospin violating contributions 
is mainly due to the poor knowledge of the
electromagnetic LECs $K_i$. This uncertainty is comparable to the
one in the isospin symmetric amplitude induced by the variations in
the strong LECs $L_i$.
\item[(3)]We have also considered the radiative process $\pi^- K^+ \to
\pi^0 K^0 \gamma$. We have explicitly demonstrated the cancellation
of the infrared divergences. The remaining radiative cross section is
strongly suppressed at threshold so that the properly redefined
(finite)  scattering length acquires no significant correction.
\item[(4)] We have also given the one--loop corrections for the
effective range $b_0$ and the P--wave scattering length $a_1$
for this channel. The isospin violating corrections are somewhat
more pronounced than for the S--wave scattering length.
\end{enumerate}
As pointed out in the introduction, for precisely predicting the
$2P-2S$ level shift in the $\pi K$ atom, one would have to extend
these considerations to the elastic scattering channel $\pi^- K^+
\to \pi^- K^+$ beyond the tree level result presented here.

\bigskip \noindent
After submission of this work, a similar analysis with comparable
results has appeared, see~\cite{Nehme}.

\bigskip \noindent
We thank G.~Ecker, A.~Nehme, and P.~Talavera for helping to resolve
an error in the first version of the manuscript.


\appendix
\section{S--wave scattering length at fourth order
\label{app:Swave}}
\noindent
In this appendix, we show the analytic expressions for the one--loop corrections
to the scattering length $a_0$ for $\pi^- K^+ \to \pi^0 K^0$. 
The isospin violating effects are expanded up to order $\epsilon$ and $e^2$.
We express all these as relative corrections $\Delta^{(4)} a_0/a_0^{(2)}$.
The low--energy constants with the infinite part subtracted are
denoted by $L_i^r$, $K_i^r$, as done conventionally.
Note however that we do not display the dependence of these various
constants on the renormalization scale $\lambda$ explicitly.
All corrections given below are of course scale independent.

\medskip \noindent
The isospin symmetric contribution is
\beqa
\frac{\Delta^{(4)}_{\rm sym} a_0}{a_0^{(2)}} &=& \frac{M_\pi^2}{F_\pi^2} \, \Biggl\{  \, 8L_5^r 
\nonumber \\ \!\!\!
-  \frac{1}{16\pi^2}  \!\!\! &&  \!\!\!\!\!\!\!\!\!\!\!
 \Biggl( \frac{1}{M_K^2 - M_\pi^2} \Biggl[ 
\biggl( 4M_K^2 - \frac{5}{2}M_\pi^2 \biggr) \log \frac{M_\pi}{\lambda}
-\frac{23}{9}M_K^2 \log \frac{M_K}{\lambda} 
+\biggl(\frac{14}{9}M_K^2 - \frac{M_\pi^2}{2}\biggr) \log \frac{M_\eta}{\lambda}
\Biggr] \nonumber\\
&+& \!\! \frac{4M_K}{9M_\pi} \Biggl[ 
\frac{\sqrt{(M_K - M_\pi)(2M_K + M_\pi)}}{M_K+M_\pi}
  \arctan \Biggl( \frac{2(M_K + M_\pi)}{M_K - 2M_\pi} \sqrt{\frac{M_K - M_\pi}{2M_K + M_\pi}} \Biggr)
\nonumber\\ && 
-\frac{\sqrt{(M_K + M_\pi)(2M_K - M_\pi)}}{M_K-M_\pi} 
  \arctan \Biggl( \frac{2(M_K - M_\pi)}{M_K + 2M_\pi} \sqrt{\frac{M_K + M_\pi}{2M_K - M_\pi}} \Biggr)
\Biggr] \Biggr) \Biggr\} 
~. 
\eeqa
\noindent
The strong isospin violating part can be expressed as
\beqa
\frac{\Delta^{(4)}_{\rm str} a_0}{a_0^{(2)}} &=& - \frac{\epsilon}{\sqrt{3}F_\pi^2}
\Biggl\{
  16  M_\pi (M_K - M_\pi) \,L_3
+ \frac{8(2M_K^4 - M_K^2M_\pi^2 - 3M_K M_\pi^3  - M_\pi^4)}{3M_K M_\pi} \, L_5^r
\nonumber \\ && \qquad \qquad
-  \frac{16(M_K^2 - M_\pi^2)(2M_K^2 + M_\pi^2)}{3M_K M_\pi} \,L_8^r
\nonumber\\
+  \,\frac{1}{16\pi^2} \!\!\!\!\!\!\!\!\!\!\! &&  \!\! \Biggl[
\frac{72M_K^5 - 56M_K^4M_\pi - 200M_K^3M_\pi^2 + 21M_K^2M_\pi^3 - 19M_KM_\pi^4 + 2M_\pi^5}
     {6M_K(M_K-M_\pi)^2(M_K+M_\pi)} 
\,M_\pi \log \frac{M_\pi}{\lambda}
\nonumber\\ &-& 
\frac{94M_K^4 - 185M_K^3M_\pi - 915M_K^2M_\pi^2 + 46M_K M_\pi^3 - 20M_\pi^4}
{27(M_K-M_\pi)^2(M_K+M_\pi)} 
\,M_\pi \log \frac{M_K}{\lambda}
\nonumber\\ &-& \Biggl(
 \frac{48M_K^7 - 48M_K^6M_\pi + 388M_K^5M_\pi^2 - 224M_K^4M_\pi^3}
  {54M_KM_\pi(M_K-M_\pi)^2(M_K+M_\pi)} 
\nonumber\\&& \qquad \qquad 
+\,\frac{192M_K^3M_\pi^4 + 259M_K^2M_\pi^5 - 269M_KM_\pi^6 - 6M_\pi^7}
  {54M_KM_\pi(M_K-M_\pi)^2(M_K+M_\pi)} \Biggr)
\log \frac{M_\eta}{\lambda}
\nonumber\\&+&
\frac{2M_\pi(16M_K^2 - 40M_KM_\pi - 47M_\pi^2)}{9(M_K-M_\pi)}
\nonumber\\&-&
\frac{2M_KM_\pi(4M_K + M_\pi)}{3(M_K+M_\pi)}
\sqrt{\frac{M_K - M_\pi}{2M_K + M_\pi}}
  \arctan \Biggl( \frac{2(M_K + M_\pi)}{M_K - 2M_\pi} \sqrt{\frac{M_K - M_\pi}{2M_K + M_\pi}} \Biggr)
\nonumber\\&-&
\frac{2M_\pi(116M_K^4 - 79M_K^3M_\pi - 156M_K^2M_\pi^2 + 74M_KM_\pi^3 + 8M_\pi^4)}
      {27(M_K-M_\pi)^2(M_K+M_\pi)} 
\nonumber\\&&\qquad
\times \sqrt{\frac{M_K + M_\pi}{2M_K - M_\pi}}
  \arctan \Biggl(\frac{2(M_K - M_\pi)}{M_K + 2M_\pi} \sqrt{\frac{M_K + M_\pi}{2M_K - M_\pi}} \Biggr)
\Biggr]\Biggr\} ~.
\eeqa
\noindent
Finally, the electromagnetic corrections at one--loop level can be written as
\beqa
\frac{\Delta^{(4)}_{\rm em} a_0}{a_0^{(2)}} & = & e^2 \, \Biggl\{ Z
\left( 8L_3-\frac{4(2M_K - M_\pi)(2M_K + 7M_\pi)}{3M_KM_\pi}\,L_5^r \right)
\nonumber\\&-& 
\frac{8}{3}(K_1^r+K_2^r) 
+ \frac{6M_K^3+2M_K^2M_\pi-6M_KM_\pi^2+M_\pi^3}{6M_K(M_K^2-M_\pi^2)}\,(2K_3^r-K_4^r)
\nonumber\\&-&
\frac{M_K(4M_K^2-7M_\pi^2)}{9M_\pi(M_K^2-M_\pi^2)} \,(K_5^r+K_6^r)
-\frac{2}{9}\,(10K_5^r+K_6^r) 
\nonumber\\&-&
\frac{M_\pi(2M_K^2+M_\pi^2)}{9M_K(M_K^2-M_\pi^2)} \,(K_9^r+K_{10}^r)
+\frac{2(4M_K^2-M_\pi^2)}{3M_KM_\pi} \,(K_{10}^r+K_{11}^r)\,\Biggr\}
\nonumber\\
+ \,\, \frac{Z e^2}{16\pi^2} \!\!\!\!\!\!\!\!\!\!\! &&  \!\! \Biggl\{
\frac{40M_K^5 - 118M_K^4M_\pi + 22M_K^3M_\pi^2 + 26M_K^2M_\pi^3 - 71M_KM_\pi^4 + 11M_\pi^5}
     {6M_K(M_K-M_\pi)^3(M_K+M_\pi)}
  \,\log \frac{M_\pi}{\lambda}
\nonumber\\ &-& 
\frac{36M_K^5 - 97M_K^4M_\pi - 211M_K^3M_\pi^2 + 126M_K^2M_\pi^3 - 400M_KM_\pi^4 + 56M_\pi^5}
     {27M_\pi(M_K-M_\pi)^3(M_K+M_\pi)}
  \,\log \frac{M_K}{\lambda}
\nonumber\\&-&
\frac{32M_K^5 + 296M_K^4M_\pi - 270M_K^3M_\pi^2 + 260M_K^2M_\pi^3 - 121M_KM_\pi^4 - 27M_\pi^5}
      {54M_K(M_K-M_\pi)^3(M_K+M_\pi)}
  \,\log \frac{M_\eta}{\lambda}
\nonumber\\ &+&
\frac{113M_K^4 - 138M_K^3M_\pi - 249M_K^2M_\pi^2 - 13M_KM_\pi^3 + 3M_\pi^4}
      {18M_K(M_K-M_\pi)^2(M_K+M_\pi)}
\nonumber\\ &-&
\frac{16M_K^2 - 17M_KM_\pi - 8M_\pi^2}
               {9(M_K + M_\pi)\sqrt{(2M_K + M_\pi)(M_K - M_\pi)}}
  \, \arctan \left(\frac{2(M_K + M_\pi)}{M_K - 2M_\pi} \sqrt{\frac{M_K - M_\pi}{2M_K + M_\pi}}\right)
\nonumber\\ &-&
\frac{32M_K^4 + 35M_K^3M_\pi - 108M_K^2M_\pi^2 - 40M_KM_\pi^3 + 44M_\pi^4}
      {27(M_K - M_\pi)^3\sqrt{(2M_K - M_\pi)(M_K + M_\pi)}}
\nonumber\\&& \qquad\qquad \times
  \arctan \left( \frac{2(M_K - M_\pi)}{M_K + 2M_\pi} \sqrt{\frac{M_K + M_\pi}{2M_K - M_\pi}} \right)
\Biggr\}
\nonumber\\
- \,\, \frac{e^2}{16\pi^2} \!\!\!\!\!\!\!\!\!\!\! &&  \!\! \Biggl\{
  \frac{12M_K^2 - 4M_KM_\pi - M_\pi^2}{2M_K(M_K+M_\pi)}
  \,\log \frac{M_\pi}{\lambda}
+ \frac{4M_K^2 + M_KM_\pi + 12M_\pi^2}{2M_\pi(M_K+M_\pi)}
  \,\log \frac{M_K}{\lambda}
\nonumber\\&&
\quad - \frac{4M_K^2 - 18M_KM_\pi - M_\pi^2}{3M_KM_\pi}
\, \Biggr\} ~.
\eeqa


\section{Infrared divergent loop diagrams\label{app:GpKg}}
The essential loop function containing an infrared divergence, stemming from
the photon exchange diagram (a) in fig.~\ref{fig:photonloops}, is
\beq
G^{\pi K\gamma}(s) = -i \int \frac{d^4k}{(2\pi)^4}
\frac{1}{\Bigl( (q_\pi\!+\!k)^2 - \Mp^2 \Bigr) 
         \Bigl( (q_K\!-\!k)^2 - \MK^2 \Bigr)
         \Bigl( k^2 - m_\gamma^2 \Bigr) } ~,
\eeq
where $s=(q_K+q_\pi)^2$. Evaluating this in the kinematical region needed
for the process in question, $s>(\MK+\Mp)^2$, we find
\beqa
{\rm Re}\,\, G^{\pi K\gamma}(s) &=&
\frac{1}{32\pi^2\sigma s} \Biggl\{
\biggl[ \log \sigma - \log\biggl(\frac{m_\gamma^2}{s}\biggr) \biggr]
\biggl[ \log\frac{1\!-\!z_1}{z_1} + \log\frac{z_2}{1\!-\!z_2} \biggr]
\\&&
+\frac{1}{2}\Bigl[ \log^2 (1\!-\!z_1) - \log^2 z_1 +\log^2 z_2 -\log^2 (1\!-\!z_2) \Bigr]
+\log\frac{z_2}{\sigma}  \log\frac{z_1}{z_2}
\nonumber\\&&
+\log\frac{1\!-\!z_1}{\sigma}\log\frac{1\!-\!z_2}{1\!-\!z_1}
-2\biggl[ {\rm Li}\biggl(\frac{\sigma}{z_2}\biggr)
         +{\rm Li}\biggl(\frac{\sigma}{1\!-\!z_1}\biggr) \biggr] 
-\frac{4\pi^2}{3} 
\Biggr\} +{\cal O}(m_\gamma) ~, \nonumber
\\
{\rm Im}\,\, G^{\pi K\gamma}(s) &=&
\frac{1}{16\pi\sigma s} \biggl\{
\log\biggl(\frac{m_\gamma^2}{s}  \biggr) - 2\log\sigma \biggr\} +{\cal O}(m_\gamma) ~,
\eeqa
where
\beqa
z_{1/2} \!&=&\! \frac{1}{2}\biggl(1-\frac{\Delta}{s}\biggr) \mp \frac{\sigma}{2} ~, \quad
\sigma \,=\, \sqrt{1-\frac{2\Sigma}{s}+\frac{\Delta^2}{s^2}} ~, \nonumber\\
\Sigma \!&=&\! \MK^2+\Mp^2 ~, \quad \Delta \,=\, \MK^2-\Mp^2 ~,
\eeqa
and 
\beq
{\rm Li}(z) = \int_1^z \frac{\log t}{1-t}dt
\eeq
is the dilogarithm or Spence function. We have checked that this loop function
coincides with the analogous one needed for $\pi\pi$--scattering as given
in \cite{Wicky,KU} when going to the limit of equal pion and kaon masses.
Other infrared divergent loop contributions stem from the wave function renormalization
of the charged pions (see e.g.~\cite{pionff}) and kaons,
\beqa
Z_{\pi^\pm} &=& -\frac{e^2}{4\pi^2} \log \frac{m_\gamma}{\Mp} + {\cal O}(m_\gamma^0) ~, \nonumber\\ 
Z_{  K^\pm} &=& -\frac{e^2}{4\pi^2} \log \frac{m_\gamma}{\MK} + {\cal O}(m_\gamma^0) ~.
\eeqa
We write the tree level amplitude for $\pi^- K^+ \to \pi^0 K^0$ as
\beq
T^{(2)} = c_0 + c_s \, s + c_t \, t + c_u \, u ~,
\eeq
where 
\beqa
c_0 \!&=&\! \frac{B}{6\sqrt{2}F^2}\biggl( (m_u\!-\!m_d)\cos\epsilon
 + (m_u\!+\!m_d\!-\!2m_s)\frac{\sin\epsilon}{\sqrt{3}} \biggr) 
 - \frac{Ze^2}{\sqrt{2}} \biggl( \cos\epsilon + \frac{\sin\epsilon}{\sqrt{3}} \biggr) ~,
\nonumber\\
c_s \!&=&\! -\frac{1}{2\sqrt{2}F^2}\biggl(\cos\epsilon + \frac{\sin\epsilon}{\sqrt{3}} \biggr) ~, \quad
c_t \,=\, \frac{\sin\epsilon}{\sqrt{6}F^2} ~, \quad
c_u \,=\, \frac{1}{2\sqrt{2}F^2}\biggl(\cos\epsilon - \frac{\sin\epsilon}{\sqrt{3}} \biggr) ~.
\eeqa
The infrared divergent piece of the real part of the one--loop amplitude $T^{(4)}$ is then given by
\beqa
\Bigl({\rm Re}\,T^{(4)} \Bigr)^{\rm div} \!\!&=&\!\! T^{(2)} \times \frac{e^2}{8\pi^2}  \biggl\{
\frac{s-\Sigma}{\sigma s} \biggl[ \log\frac{1\!-\!z_1}{z_1} + \log\frac{z_2}{1\!-\!z_2} \biggr]
 \log\frac{m_\gamma}{\sqrt{s}}
-\log\frac{m_\gamma}{\Mp} -\log\frac{m_\gamma}{\MK} \biggr\} 
\nonumber\\
&\equiv&\!\! T^{(2)} \times \frac{e^2}{8\pi^2} \,L(m_\gamma) ~.
\eeqa
The total cross section is obtained from the amplitude via
\beq
\sigma = \frac{1}{64\pi^2s}\frac{|\qout|}{|\qin|} \int |T|^2 d\Omega ~,
\label{totalXsec}
\eeq
with $|\qin|$ and $|\qout|$ defined as in eqs.~(\ref{qin}), (\ref{qout}).
The infrared divergent part of the total cross section can be calculated from 
eq.~(\ref{totalXsec}) to be
\beqa
\sigma^{\rm div} &=& \frac{e^2}{(4\pi)^3} \frac{|\qout|}{|\qin|s}
\Biggl\{ \biggl[ c_0 + \biggl( c_s - \frac{c_t+c_u}{2}\biggr) s
+ \frac{c_t+c_u}{2} \Bigl( \MK^2 \!+\! \MKn^2 \!+\! \Mp^2 \!+\! \Mn^2 \Bigr)
\nonumber\\&& 
- \frac{c_t-c_u}{2} \frac{(\MK^2 \!-\! \Mp^2)(\MKn^2 \!-\! \Mn^2)}{s} \biggl]^2
+\frac{4}{3}(c_t-c_u)^2 |\qin|^2 |\qout|^2 \Biggr\} L(m_\gamma) ~.
\label{sigdiv}
\eeqa


\section{The radiative cross section\label{app:radXsec}}
\begin{figure}[htb]
\centerline{
\epsfysize=4.cm
\epsffile{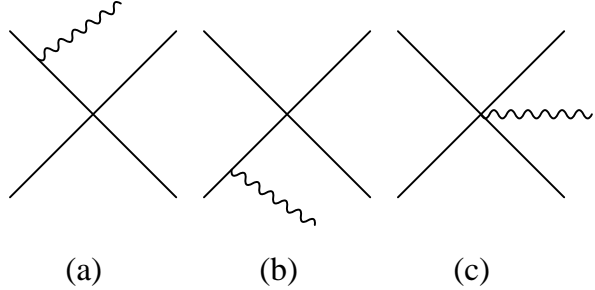}
}
\caption{Diagrams contributing to $\pi^- K^+ \to \pi^0 K^0 \gamma$ at tree level.
\label{fig:radiation}
}
\end{figure}
\noindent
The amplitude for the process
\beq
\pi^-(q_\pi)\, K^+(q_K) \to \pi^0(p_\pi)\, K^0(p_K)\, \gamma(l)
\eeq
is given, to lowest order, by
\beqa
T^\gamma &=& 
\frac{e\, \epsilon^\ast \cdot (2q_K-l)}{2q_K\cdot l - m_\gamma^2}
 \Bigl( c_0 + c_s\, s_2 + c_t\, t_\pi +c_u\, u_\pi \Bigr) 
\nonumber \\ &-& 
\frac{e\, \epsilon^\ast \cdot (2q_\pi-l)}{2q_\pi\cdot l - m_\gamma^2}
 \Bigl( c_0 + c_s\, s_2 + c_t\, t_K +c_u\, u_K \Bigr)
\nonumber \\ &+&
\frac{e}{2\sqrt{2}F^2}\, \epsilon^\ast \cdot \biggl\{
\Bigl(q_K-q_\pi+p_K-p_\pi\Bigr) \cos\epsilon
+ \Bigl(q_K-q_\pi-3p_K+3p_\pi\Bigr) \frac{\sin\epsilon}{\sqrt{3}} \biggr\} ~,
\label{ampgamma}
\eeqa
where
\beqa
s_2 = (p_\pi+p_K)^2 ~, \quad 
t_{\pi/K} = (q_{\pi/K}-p_{\pi/K})^2 ~, \quad
u_{\pi/K} = (q_{\pi/K}-p_{K/\pi})^2 ~.
\eeqa
The three terms in eq.~(\ref{ampgamma}) correspond to the three diagrams
in fig.~\ref{fig:radiation}.
The total cross section can be calculated from this amplitude by
\beq
\sigma^\gamma = \frac{1}{(2\pi)^5} \frac{1}{4 |\qin| \sqrt{s}} 
\int dE_l \int d\Omega_{p_\pi} \int d\Omega_{p_\pi l}  
\frac{|{\bf p_\pi}| |{\bf l}|}{8\left( E_{K^0}+E_{\pi^0}\left(1+
 \frac{|{\bf l}|}{|{\bf p_\pi}|}\cos \theta_{p_\pi l} \right)\right)} |T^\gamma|^2 ~,
\label{sigmagamma}
\eeq
where $\theta_{p_\pi}$ refers to the the angle between the outgoing pion and the
axis of the incoming particles (in their center--of--mass system),
and $\theta_{p_\pi l}$ denotes the relative angle between the outgoing pion and
the photon.


\subsection{Infrared divergence\label{app:radXsec:div}}
One important check resulting from the exact calculation of
the radiative cross section is that the 
infrared divergences indeed cancel when combining radiative and non--radiative cross sections,
we therefore demonstrate how to isolate the infrared divergent parts in 
the cross section eq.~(\ref{sigmagamma}).
These stem exclusively
from the lower integration bound of the $E_l$--integration, therefore it suffices
to collect the leading powers in $1/E_l$ of the integrand. 
The squared matrix element simplifies considerably in this approximation, one may set
$s_2 \to s$, $t_{\pi/K} \to t$, $u_{\pi/K} \to u$, and obtains
\beq
|T^\gamma|^2 = e^2 \biggl\{ -\frac{\Mp^2}{(l\cdot q_\pi)^2} -\frac{\MK^2}{(l\cdot q_K)^2}
+ \frac{s-\Mp^2-\MK^2}{l\cdot q_\pi \,\, l\cdot q_K} \biggr\}
\Bigl( c_0 + c_s s+c_t t+c_u u \Bigr)^2 +{\cal O}(E_l^{-1}) ~.
\eeq
As is well known from Quantum Electrodynamics, infrared divergences arise only from diagrams with
soft photon radiation \emph{from external legs} (diagrams (a), (b) in 
fig.~\ref{fig:radiation}), and indeed contributions
from the diagram where the photon is radiated from the $\pi K$--vertex
(diagram (c) in fig.~\ref{fig:radiation}) play no role here.
Furthermore, we can set $|{\bf p_\pi}| = |\qout|$ (which is, as above, the modulus
of the outgoing momentum of the \emph{non--radiative} process), and use the abbreviations 
$y=\cos \theta_{p_\pi l}$, $z=\cos \theta_{p_\pi}$, to obtain 
\beqa
\sigma^\gamma \!&=&\! \frac{e^2}{(4\pi)^3} \frac{|\qout|}{4 |\qin| s}
\int_{m_\gamma}^{E_\gamma^{\rm max}}\!\! dE_l\, E_l \int_{-1}^1 dz \int_{-1}^1 dy  
\int_0^{2\pi} \frac{d \phi_{p_\pi l}}{2\pi}
\nonumber \\ \!& \times &\!
\biggl\{ -\frac{\Mp^2}{(l\cdot q_\pi)^2} -\frac{\MK^2}{(l\cdot q_K)^2}
+ \frac{s-\Mp^2-\MK^2}{l\cdot q_\pi \,\, l\cdot q_K} \biggr\}
\Bigl( c_0 + c_s s+c_t t+c_u u \Bigr)^2 +{\cal O}(m_\gamma^0) ~.\qquad
\label{Xsec:div}
\eeqa
Note that in this approximation, the term $c_0 + c_s s+c_t t+c_u u$ depends, 
of all integration variables, only on $z$ (via $t$ and $u$). 
The $\phi_{p_\pi l}$ integration is relatively straightforward when using
\beq
l\cdot q_{\pi/K} = E_l\, |\qin|\, \Bigl( S_{\pi/K} \mp \cos \alpha \Bigr) ~,
\eeq
where
\beq
S_{\pi/K} = \frac{s \mp \MK^2 \pm \Mp^2}{2 |\qin| \sqrt{s}} ~, \quad
\cos \alpha = y z -\sqrt{1-\smash{y}^2} \sqrt{1-z^2} \cos \phi_{p_\pi l} ~~,
\eeq
such that, for example, the first term in the curly brackets in eq.~(\ref{Xsec:div}) yields
\beq
\Mp^2 \int_{-1}^1 dy  
\int_0^{2\pi} \frac{d \phi_{p_\pi l}}{2\pi}
\frac{1}{(l\cdot q_\pi)^2} 
= \frac{\Mp^2}{E_l^2 |\qin|^2} \int_{-1}^1 dy 
\frac{S_\pi -y z}{\bigl(S_\pi^2-1-2S_\pi yz +y^2+z^2\bigr)^{3/2}}
=\frac{2}{E_l^2} ~,
\eeq
and the other terms can be evaluated similarly. The only remaining $z$--dependence
is the one which is also inherent in the integration of the non--radiative
cross section, 
\beq
\int_{-1}^1 dz   \Bigl( c_0 + c_s s + c_t t+c_u u \Bigr)^2 ~.
\eeq
Altogether, one ends up with
\beqa
\sigma^\gamma &=& \frac{e^2}{(4\pi)^3} \frac{|\qout|}{|\qin| s}
\log \biggl( \frac{E_\gamma^{\rm max}}{m_\gamma} \biggr)
\Biggl\{ \frac{s-\MK^2-\Mp^2}{2|\qin|\sqrt{s}} \biggl(
\log\frac{S_K+1}{S_K-1} + \log\frac{S_\pi+1}{S_\pi-1} \biggr) -2 \Biggr\}
\nonumber \\ && \times
\Biggl\{ \biggl[ c_0 + \biggl( c_s - \frac{c_t+c_u}{2}\biggr) s 
+ \frac{c_t+c_u}{2} \Bigl( \MK^2 \!+\! \MKn^2 \!+\! \Mp^2 \!+\! \Mn^2 \Bigr) 
\\ && \qquad 
- \frac{c_t-c_u}{2} \frac{(\MK^2 \!-\! \Mp^2)(\MKn^2 \!-\! \Mn^2)}{s} \biggl]^2
+\frac{4}{3}(c_t-c_u)^2 |\qin|^2 |\qout|^2 \Biggr\} +{\cal O}(m_\gamma^0) ~, 
\nonumber
\label{siggammadiv}
\eeqa
which cancels the $m_\gamma$--divergence in eq.~(\ref{sigdiv}).


\subsection{The radiative cross section at threshold\label{app:radXsec:thr}}
The other aspect of the radiative cross section on which one would like to have
analytical information is the remainder at threshold. 
As the incoming charged particles are at rest at threshold, the angular integration
for the total cross section simplifies considerably:
the integration $d\Omega_{p_\pi}$ is trivial, and so is the integration $d\phi_{p_\pi l}$, 
hence the only angular variable to integrate is $dy=d\cos\theta_{p_\pi l}$, 
which is most conveniently done in the center--of--mass frame of the outgoing
$\pi^0 K^0$ system.
Large cancellations take place, such that one obtains a surprisingly simple expression:
\beqa
\int_{-1}^1 dy \, |T^\gamma|^2_{\rm thr}
\!&=&\! \frac{e^2}{24F^4} \Bigl(\cos\epsilon -\sqrt{3}\sin\epsilon \Bigr)^2
\frac{(\MK\!+\!\Mp)^2}{\MK\Mp} 
\nonumber\\ \!&\times&\!
\frac{\Bigl(s_2-(\MK\!-\!\Mp)^2\Bigr)\Bigl(s_2-(\MKn\!-\!\Mn)^2\Bigr)
      \Bigl(s_2-(\MKn\!+\!\Mn)^2\Bigr)}{s_2^2} ~.\qquad
\label{Athr_intz}
\eeqa
In order to obtain eq.~(\ref{Athr_intz}), it is essential to take the radiation from the four--meson
vertex (diagram (c) in fig.~\ref{fig:radiation})
into account, the vertex of which is linked to the $\pi^-K^+ \to \pi^0 K^0$ vertex
by gauge symmetry. In fact, we find that just by expressing the coefficients for the 5--point vertex 
(the third term in eq.~(\ref{ampgamma})) by
$c_t$ and $c_u$, any dependence on $c_0$ and $c_s$ drops out completely.
This is why for the case of $\pi^+ \pi^- \to \pi^0 \pi^0 \gamma$, the cross section 
vanishes at threshold: the lowest--order $\pi^+ \pi^- \to \pi^0 \pi^0$ vertex 
is proportional to $s-\Mn^2$ (in the $\sigma$ gauge), and the 
$\pi^+ \pi^- \to \pi^0 \pi^0 \gamma$ point vertex vanishes (again in the $\sigma$ gauge).
Furthermore, the result of eq.~(\ref{Athr_intz}) is proportional to $(c_t-c_u)^2$ which 
demonstrates that only the particular ``asymmetry'' of the reaction in question
allows a finite remainder to survive at threshold.

\medskip \noindent
Transforming the remaining integration over the energy of the outgoing photon into an integration
of $s_2$, we obtain the total cross section at threshold via
\beqa
\sigma^\gamma_{\rm thr} &=&
\frac{1}{(8\pi)^3} \frac{|\qout|}{|\qin|s} \int_{s_2^{\rm thr}}^s ds_2
\biggl( \frac{s-s_2}{s_2} \biggr) 
\sqrt{\frac{\bigl(s_2-(\MKn\!-\!\Mn)^2\bigr)\bigl(s_2-(\MKn\!+\!\Mn)^2\bigr)}
           {\bigl(s  -(\MKn\!-\!\Mn)^2\bigr)\bigl(s  -(\MKn\!+\!\Mn)^2\bigr)}}
\nonumber\\&&\qquad\qquad\qquad
\times \int_{-1}^1 dy \, |T^\gamma|^2_{\rm thr} ~~,
\label{sigthr}
\eeqa
where $s_2^{\rm thr} = (\MKn+\Mn)^2$ is the minimal energy (squared)
for the final $\pi^0 K^0$ system.
Eq.~(\ref{sigthr}) is evaluated to leading order in isospin violation parameters, 
yielding the result
\beqa
\sigma^\gamma_{\rm thr} &=&
\frac{e^2}{105(2\pi)^3 F_\pi^4} \frac{|\qout|}{|\qin|s} \frac{\MK\Mp}{\MK\!+\!\Mp} 
\Bigl( \MK \!-\! \MKn \!+\! \Mp \!-\! \Mn \Bigr)^3 + {\cal O}(\delta^5) ~.
\label{sigapprox}
\eeqa
In terms of fundamental isospin breaking parameters, the meson mass difference 
in eq.~(\ref{sigapprox}) can be expressed as
\beqa
\MK - \MKn + \Mp - \Mn &=& 
\frac{B (m_u\!-\!m_d)}{2M_K}
+Ze^2F^2\biggl(\frac{1}{M_K}+\frac{1}{M_\pi}\biggr) + {\cal O}(\delta^2) ~.
\eeqa


\end{document}